\documentclass[runningheads]{llncs}

\usepackage{graphicx}

\usepackage{array}
\usepackage{multirow}
\usepackage{xspace}
\usepackage{color,soul}
\usepackage{adjustbox}
\usepackage{comment}
\usepackage{url}
\usepackage[utf8]{inputenc}
\usepackage{algorithm}
\usepackage{algorithmic}
\usepackage{float}
\usepackage{enumitem}
\usepackage{amsmath}

\newcommand{\method}{SHIELD\xspace}

\usepackage{color}

\begin{document}
\title{\method: APT Detection and Intelligent Explanation Using LLM}
\titlerunning{\method: APT Detection and Intelligent Explanation Using LLM}
%
\author{Parth Atulbhai Gandhi \and Prasanna N. Wudali \and Yonatan Amaru \and 
Yuval Elovici \and
Asaf Shabtai}
\authorrunning{Gandhi et al.}
%
\institute{Ben-Gurion University of the Negev}
\maketitle

\begin{abstract}
Advanced persistent threats (APTs) are sophisticated cyber attacks that can remain undetected for extended periods, making their mitigation particularly challenging. 
Given their persistence, significant effort is required to detect them and respond effectively. 
Existing provenance-based attack detection methods often lack interpretability and suffer from high false positive rates, while investigation approaches are either supervised or limited to known attacks. 
To address these challenges, we introduce \method, a novel approach that combines statistical anomaly detection and graph-based analysis with the contextual analysis capabilities of large language models (LLMs). 
\method leverages the implicit knowledge of LLMs to uncover hidden attack patterns in provenance data, while reducing false positives and providing clear, interpretable attack descriptions. 
This reduces analysts' alert fatigue and makes it easier for them to understand the threat landscape. 
Our extensive evaluation demonstrates \method's effectiveness and computational efficiency in real-world scenarios. 
\method was shown to outperform state-of-the-art methods, achieving higher precision and recall. 
\method's integration of anomaly detection, LLM-driven contextual analysis, and advanced graph-based correlation establishes a new benchmark for APT detection.


\end{abstract}

\section{Introduction}

Advanced persistent threats (APTs) are becoming a growing concern for today's organizations due to their sophistication and stealth. 
The 2020 SolarWinds supply chain attack~\cite{lazarovitz2021deconstructing} in which attackers remained undetected for over nine months while compromising 18,000 organizations and major U.S. government agencies demonstrates the harm APTs can cause.
Another APT, APT-C-36 (Blind Eagle),\footnote{\small\url{https://www.trendmicro.com/en_us/research/21/i/apt-c-36-updates-its-long-term-spam-campaign-against-south-ameri.html}} has been actively targeting the financial and government sectors in South America since 2018, employing advanced social engineering and fileless malware techniques to evade detection while maintaining a long-term presence in compromised networks. These incidents, which show how APTs can persist undetected for long periods of time, highlight the urgent need for sophisticated detection approaches that can identify threats without overwhelming security analysts with false positives.

The analysis of system provenance data, which captures detailed relationships and interactions between processes, files, and network connections over time, has emerged as a promising approach for APT detection~\cite{zipperle2022provenance,inam2023sok}. 
By maintaining a comprehensive record of system activities and their dependencies, provenance data can expose subtle patterns and lateral movements characteristic of sophisticated attacks. 
Existing detection methods relied on expert-defined rules 
to identify known attack patterns in provenance data~\cite{hossain2017sleuth,milajerdi2019holmes,hassan2020tactical,kurniawan2022krystal}. 
While effective against known threats, these rule-based approaches require constant updates and struggle with zero-day attacks. 
To address these limitations, researchers shifted toward machine learning-based anomaly detection methods that model normal system behavior to identify deviations. These approaches range from statistical analysis~\cite{hossain2017sleuth,hassan2019nodoze,wang2020you,kurniawan2022krystal,dong2023distdet,Li_2024} to deep learning models using path~\cite{du2017deeplog,zhang2019robust,guo2021logbert,alsaheel2021atlas} and graph-based architectures~\cite{han2020unicorn,wang2022threatrace,zengy2022shadewatcher,jia2023magic,yang2023prographer} for pattern recognition.

However, current machine learning (ML) methods face two significant limitations. 
These methods require extensive training data and struggle to adapt to the dynamic nature of both system behaviors and evolving attack patterns, leading to a high rate of false positives. 
In addition, existing approaches for interpreting alerts rely on labeled training data, which are tedious to generate and quickly become outdated~\cite{alsaheel2021atlas,milajerdi2019holmes,hossain2017sleuth,hassan2020tactical}, or focus on ML explainability rather than the specific characteristics of attacks.

Recent advances in large language models (LLMs) have introduced new capabilities for understanding and reasoning about complex system behaviors. 
These models, which have a remarkable ability to adapt to different contexts and provide logical explanations for their findings, could serve as the basis of new APT detection approaches, overcoming the shortcomings of existing approaches. 
However, LLMs also have two critical limitations. 
First, directly feeding raw system logs to LLMs is impractical and inefficient, requiring to provide them with pre-processed logs using traditional security analysis methods like anomaly detection. 
Second, LLMs are prone to hallucination, which can be addressed through multiple complementary techniques like prompt engineering, chain-of-thought reasoning, and validation checks.

To address these challenges, we propose \method, a novel framework that combines unsupervised anomaly detection with language model reasoning for APT detection and investigation. 
Our framework accumulates system logs in a sliding window mechanism for analysis, while leveraging a temporal correlation engine to track the historical evolution of suspicious events to enable the detection of attacks spanning over a long period of time. 
The proposed short-window mechanism is critical for rapid threat identification, as it detects and responds to attacks with minimal latency. 
\method (1) employs statistical analysis to flag anomalous events; (2) constructs provenance graphs to determine relationships; (3) prunes benign activities; and (4) clusters suspicious events using community detection algorithms. 
These steps reduce large amounts of system events to concise groups of suspicious activities, preserving significant attack patterns. 
An LLM then performs multi-stage reasoning on the identified clusters, providing interpretable insights and mapping them to the APT kill chain. 

The temporal correlation engine continuously compares new alerts with historical events. 
It employs dynamic confidence scoring using decay and reinforcement mechanisms. 
The decay mechanism lowers the confidence score if flagged processes consistently exhibit benign behavior, while the reinforcement mechanism increases scores as additional attack stages are detected across time windows. 
This dual-mechanism approach maintains a comprehensive view of slow-evolving attacks while preserving real-time detection capabilities and ensuring low false positive rates. 
Unlike existing approaches that focus solely on anomaly or rule-based detection, \method combines these techniques with LLM-based reasoning to bridge the gap between detection and actionable security insights.

We evaluate \method on four diverse datasets: the DARPA Eng. 3 CADETS and THEIA datasets,\footnote{\scriptsize\url{https://github.com/darpa-
i2o/Transparent-Computing}} the Public Arena dataset\footnote{\scriptsize\url{https://github.com/security0528/PublicArena}}, and an in-house dataset containing the Blind Eagle APT-C-36 campaign. 
Using only 30\% of each dataset for training, \method achieved perfect precision on the CADETS dataset and maintained high recall (0.93-1.00) across all datasets. 
\method successfully tracked attacks spanning multiple days, with detailed attack narratives mapping precisely to APT stages from initial compromise through lateral movement and data exfiltration. 
Most notably, when analyzing the key CADETS attack sequence, \method identified 25 true positive events with zero false positives, while baseline methods generated over 4000 false events requiring analyst investigation.

The key contributions of this work can be summarized as follows:
(1) To the best of our knowledge, we are the first to use LLMs for APT detection and investigation. 
By integrating LLM reasoning with traditional security analysis, our framework achieves high detection accuracy while minimizing false alarms.
(2) We developed a novel engine that can track attacks spanning over a long time period using dynamic confidence scoring with reinforcement and decay mechanisms, effectively identifying slow-evolving, stealth attacks.   
(3) We designed a novel framework capable of generating human-interpretable, comprehensive attack summaries that maps attack events to kill-chain stages. 
These summaries, enriched with relevant IoCs, enable security teams to quickly analyze and respond to potential threats.
(4) We leverage the capabilities of off-the-shelf LLMs, eliminating the need for traditional model training. 
Unlike deep learning-based methods, that suffer from concept drift, our system maintains adaptability through prompt-based contextualization of organizational changes.
    
\section{\label{sec:related}Related Work}

Provenance-based approaches for APT detection have been developed to address increasingly sophisticated cybersecurity challenges. 
Rule-based techniques~\cite{milajerdi2019holmes,yu2019needle} utilize predefined security policies and heuristic rules to identify attack patterns. 
While these methods offer detailed, fine-grained event-level detection~\cite{inam2023sok} and typically maintain low false positive rates, they require extensive manual efforts and are less effective in detecting zero-day exploits~\cite{hossain2020combating}. \\
\indent Anomaly-based detection methods, including statistical techniques~\cite{hossain2017sleuth,hassan2019nodoze,wang2020you,kurniawan2022krystal,dong2023distdet,Li_2024}, path-based approaches~\cite{du2017deeplog,zhang2019robust,guo2021logbert,alsaheel2021atlas}, and graph-based methods~\cite{han2020unicorn,wang2022threatrace,zengy2022shadewatcher,jia2023magic,yang2023prographer}, have also demonstrated promising results in detecting APTs. 
However, these methods often struggle to adapt to changes in system behavior over time, which can lead to high false positive rates overwhelming security analysts. 
In addition, the alerts generated by these techniques often lack interpretability, making it challenging for security analysts to understand and investigate the anomalies detected. \\
\indent Moreover, some of the recent studies~\cite{alsaheel2021atlas,wang2022threatrace,zengy2022shadewatcher,jia2023magic,han2020unicorn,Li_2024} have two methodological limitations. 
First, these methods heavily depend on large training datasets while utilizing disproportionately small testing windows, a practice that conflicts with the dynamic and ever-evolving nature of real-world data environments. 
Secondly, some studies have identified instances of data leakage~\cite{wang2022threatrace,jia2023magic}, where models were unintentionally trained on future data, potentially leading to inflated performance metrics. 
These limitations highlight the need for additional research in order to develop new approaches to enhance the effectiveness and robustness of these approaches, thereby promoting their wider adoption and practical implementation
~\cite{van2019sok,dong2023we}. \\
\indent Other recent research focused on using transformers and LLMs for anomaly detection in logs. 
For instance, LAnoBERT~\cite{lee} employs regular expressions for minimal log preprocessing, enhancing flexibility across different formats; LogGPT~\cite{han2023loggpt} uses a GPT-2 model with sequential prediction and a top-K reward mechanism to improve detection capabilities; and LogPrompt~\cite{liu2024} leverages advanced prompt strategies to boost LLM performance in log analysis. 
While these approaches enable advanced log analysis, research on their application on system-level provenance entities and use in identifying APT attacks remains limited. \\
\indent The main limitations identified in the proposed methods and prior studies can be summarized as follows: 
(1) rule-based methods require manual efforts and are unable to detect zero-day attacks;
(2) anomaly-based methods fail to adapt to evolving system behavior over time, resulting in high false positive rates; 
(3) alerts generated by anomaly-based methods lack interpretability, making it difficult for security analysts to investigate detected anomalies;
(4) in many cases, the evaluation performed used training data that do not reflect real-world scenarios, leading to inflated (incorrect) performance results; 
(5) existing methods face challenges with scalability and have high computational demands, limiting their practical deployment; and
(6) insufficient application of LLMs for the tasks of APT detection and investigation, particularly with the use of provenance data.
In light of these gaps and limitations, we propose \method, a novel framework designed to address the challenges of APT detection and investigation in provenance-based systems.

\section{\label{sec:method}Proposed Method}

\begin{figure}[ht]
\setlength{\abovecaptionskip}{3pt}
\setlength{\belowcaptionskip}{0pt}
\centering
\includegraphics[width=\textwidth]{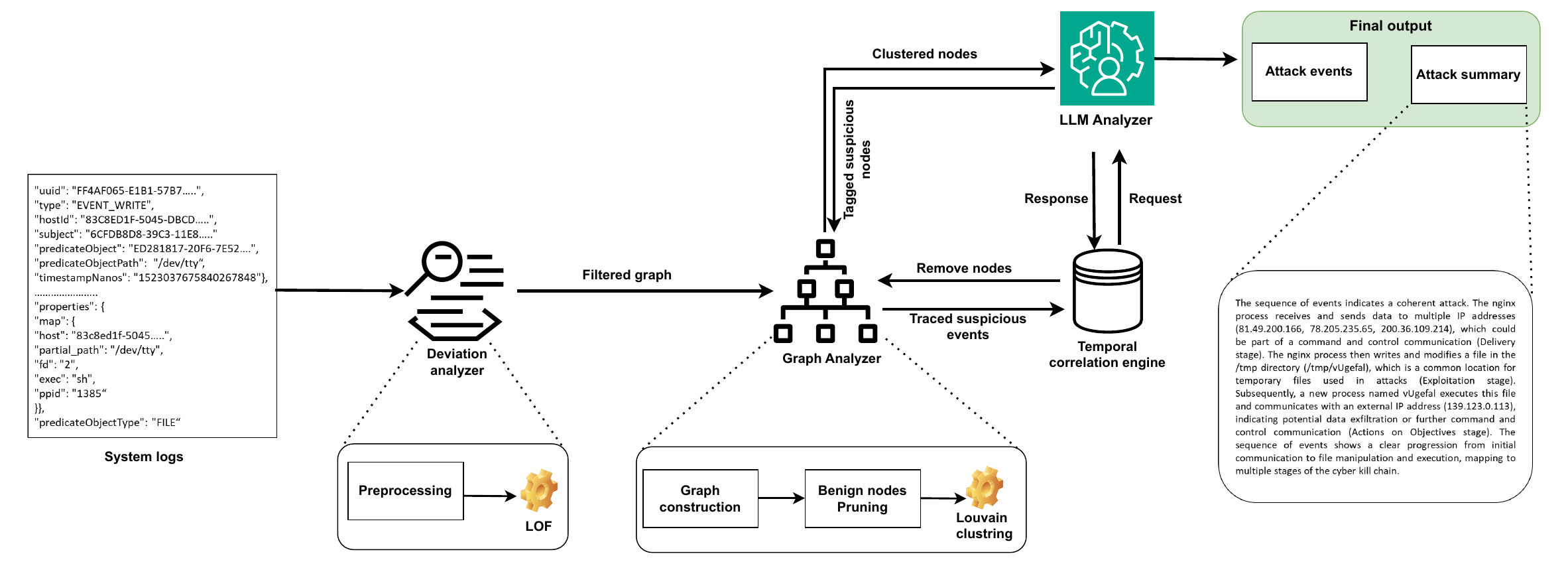}
\caption{Overview of the \method pipeline, with an example demonstrating the interaction between the four modules.}
\label{fig:pipeline}
\end{figure}

\method consists of four interdependent modules working in a iterative feedback-loop (Fig~\ref{fig:pipeline}). 
The deviation analyzer examines system logs to identify anomalous events. 
The graph analyzer processes the anomalous events, constructs a graph, prunes benign nodes, and clusters suspicious nodes into communities. 
The LLM analyzer evaluates these communities, identifying attack patterns and assigning confidence scores. 
Finally, the temporal correlation engine preserves historical attack data, correlates events across time, and re-initiates analysis when necessary, ensuring continuous and adaptive threat detection.

\begin{figure}[!t]
\centering
\includegraphics[width=1.2\textwidth]{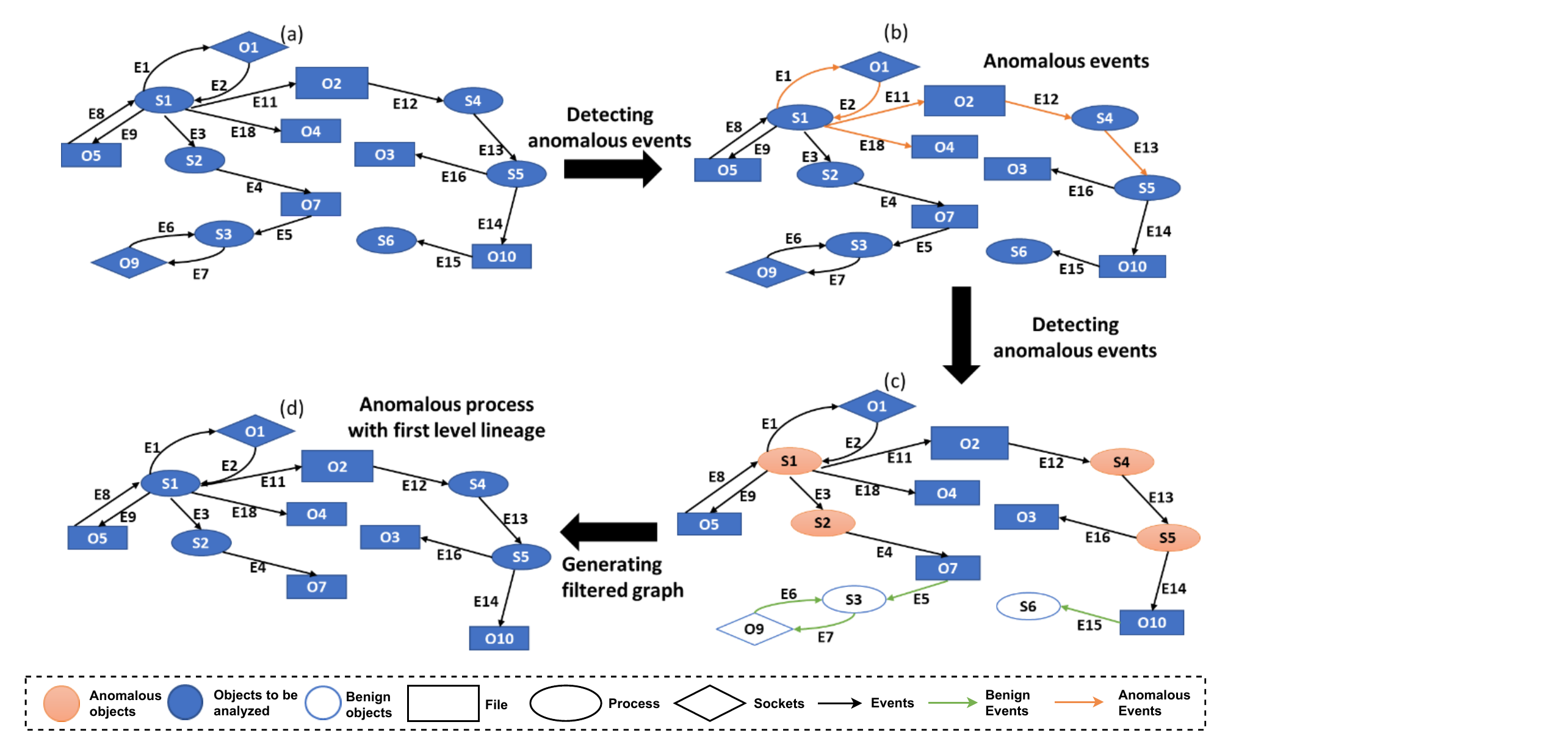}
\caption{Deviation analyzer: Detection of event-level anomalies, followed by the addition of the processes and their first-level ancestors and descendants for further analysis.}
\label{fig:deviation_analyzer}
\end{figure}

\noindent\textbf{Deviation Analyzer. }
The deviation analyzer serves as the foundation of \method's detection capabilities by identifying behavioral deviations in system logs (see example in Fig.~\ref{fig:deviation_analyzer}). 
It uses a local outlier factor (LOF) model provided with baseline system activities as a reference.
This model processes raw logs to produce a filtered graph of anomalous processes and their immediate lineage. 

\noindent\textit{(a) Log structure and notation.} The system logs capture detailed process-object interactions in the operating system. 
Each log entry $l_i$ in our dataset is represented as: 
  $l_i = (p_i, n_i, e_i, o_i, d_i, t_i)$,
where $p_i$ represents the process identifiers, $n_i$ the process name, $e_i$ the event type, $o_i$ the object identifiers, $d_i$ the object data, and $t_i$ the timestamp indicating time at which the event occurred. 
To efficiently process these logs, deviation analyzer employs a mapping mechanism that converts categorical variables into numerical representations while maintaining a bidirectional mapping for result interpretation.

\noindent\textit{(b) Outlier detection.} The anomalous event detection task involves several key steps. 
First, we preprocess the system logs by removing duplicate entries to ensure that unique process-object interactions are analyzed. 
The features used for anomaly detection include the numerical representations of the logs $\langle p_i, e_i, o_i \rangle$, which are standardized using \texttt{StandardScaler} to ensure uniform feature scaling.
\begin{equation}
A(l_i) =
\begin{cases}
1 & \text{if } \text{LOF}_k(l_i) > \tau, \\
0 & \text{otherwise}
\end{cases}
\end{equation}
where $\tau$ is the contamination threshold set at 0.1 and $k$ is the number of neighbors set at 20 used for density estimation by the LOF algorithm; these values were set based on an empirical study we performed.

\noindent\textit{(c) Filtered graph creation.} Based on the detected anomalous events (red arrows in Fig.~\ref{fig:deviation_analyzer}b), the deviation analyzer identifies the processes performing these events (red ovals in Fig.~\ref{fig:deviation_analyzer}c). 
Then, for each anomalous process identified, the deviation analyzer extracts: (1) the process's direct ancestors through fork events, which are identified by matching the object identifier $(o_i)$ with the anomalous process's identifier $(p_i)$, and (2) the process's immediate descendants, found by examining fork events where the anomalous process is the parent, as shown in Fig.~\ref{fig:deviation_analyzer}c (process $S2$).
Choosing one-hop lineage stems from a well-known property of system-level provenance graphs where a process is responsible for transferring data between two objects~\cite{inam2023sok}. 
This lineage tracking is formalized as:
\begin{equation}
R = \bigcup_{i} G_i \quad \forall l_i : A(l_i) = 1
\end{equation}
where $G_i$ represents the subgraph containing the anomalous process and its one-hop lineage as depicted in Fig.~\ref{fig:deviation_analyzer}d. \\

\begin{figure}[ht]
\setlength{\abovecaptionskip}{3pt}
\setlength{\belowcaptionskip}{0pt}
\centering
\includegraphics[width=\textwidth]{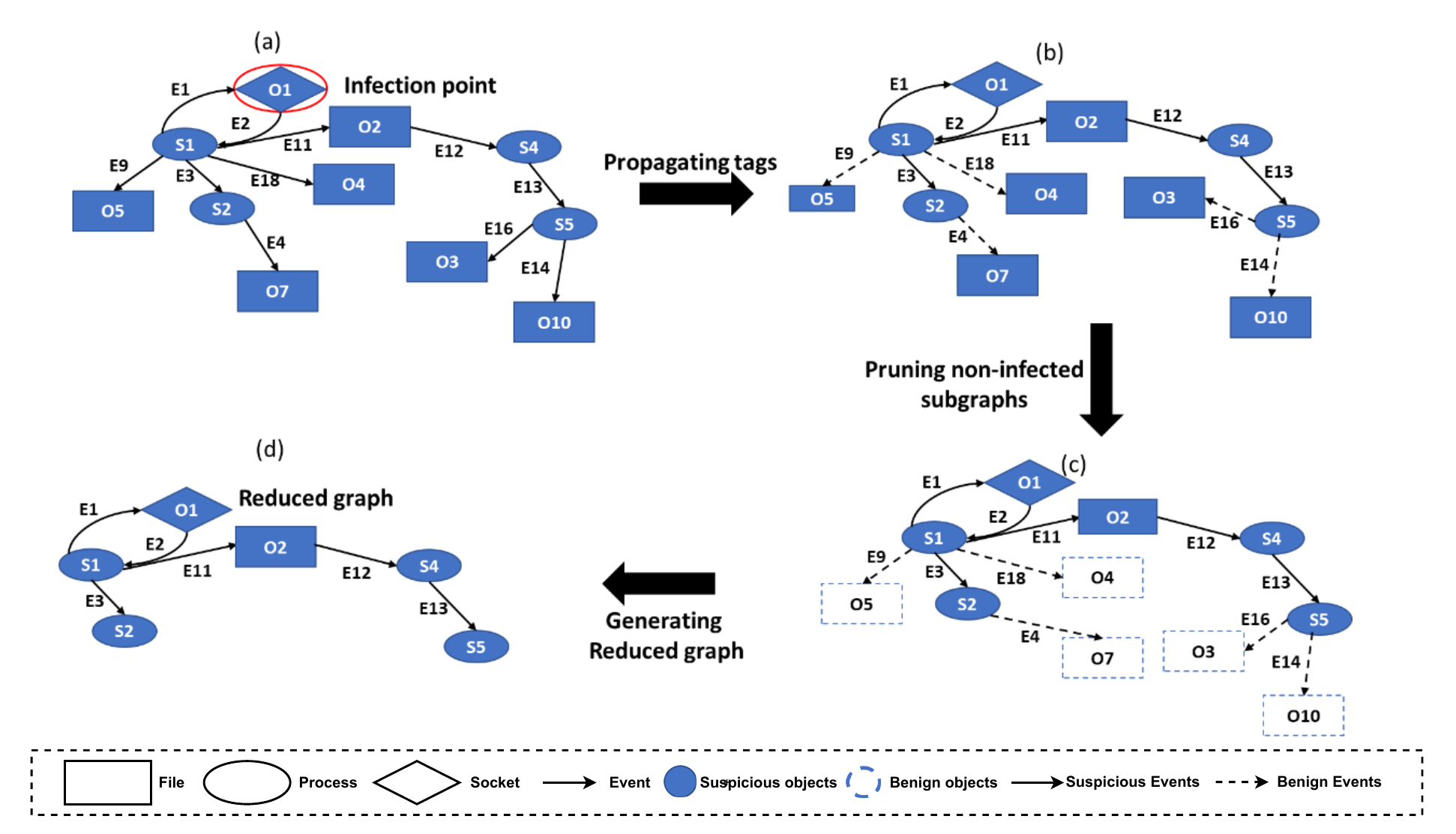}
\caption{Graph analyzer: Detection of infection points, followed by tag-propagation and iterative pruning of benign entities, resulting in a reduced graph structure optimized for further analysis.}
\label{fig:graph_analyzer}
\end{figure}

\noindent\textbf{Graph Analyzer. }
Using the filtered graph from the deviation analyzer as input, the graph analyzer examines relationships between processes, identifying potential points of infection, prunes benign nodes, and clusters the suspicious nodes using community detection algorithms (as illustrated in Fig.~\ref{fig:graph_analyzer}).

\noindent\textit{(a) Detection of initial infection points.} 
\method focuses on attacks originating from entities through network communication interfaces (sockets), which serve as the connection points between the victim's system and external entities (e.g., a user visiting a malicious website that downloads a payload to the system).
As illustrated in Fig~\ref{fig:graph_analyzer}, the socket is the object from which the attack was initiated. 
We refer to all such external sockets as initial infection points. 
In the example presented in Fig.~\ref{fig:graph_analyzer}a, socket object \textit{O1} is the initial infection point.

\noindent\textit{(b) Suspicious tag propagation.} The graph analyzer propagates the suspicious tags from the initial infection points to other nodes in the provenance graph; (initially, all of the processes that receive data from the initial infection points are tagged as suspicious.) 
Then, these tags are propagated to the nodes that receive data from these suspicious processes. 
Tags are thus propagated to all the entities that exist in the path. 
A tag propagation path only ends when: (1) a socket object is encountered, or (2) entities other than socket objects do not relay the data they receive.
Fig.~\ref{fig:graph_analyzer}b shows an example of tag propagation paths in the graph.
To formalize this propagation process, we define a tag propagation function $T: V \rightarrow {0,1}$ that marks vertices as suspicious:

\begin{equation}
T(v) =
\begin{cases}
1 & \text{if } v \in I \lor \left(\exists u \in V : (u,v) \in E \land \text{relaysData}(v)\right), \\
0 & \text{otherwise}
\end{cases}
\end{equation}

where $relaysData(v)$ is a Boolean function indicating whether vertex $v$ forwards its received data to other nodes, and $I$ is the set of all initial infection points.

\noindent\textit{(c) Pruning non-infected subgraphs.} After the tags have been propagated, the graph analyzer prunes all nodes that are not tagged and are therefore considered unimportant (illustrated in Fig.\ref{fig:graph_analyzer}c in which entities O3, O4, O5, O7, and O10 are removed).
Based on this pruning criteria, we can formally define the reduced graph $G_R = (V_R, E_R)$ as:

\begin{equation}
\begin{aligned}
V_R &= \{ v \in V \mid T(v) = 1\}, \\
E_R &= \{ (u, v) \in E \mid u, v \in V_R \}.
\end{aligned}
\end{equation}

\noindent\textit{(d) Community detection.} The Louvain algorithm is applied to the reduced graph, clustering related nodes into communities to identify groups of processes that may be working together in coordinated attack activities. We chose this algorithm since: (1) it efficiently handles large, high-dimensional graphs, (2) it is known for producing high-modularity partitions, and (3) prior studies have demonstrated its effectiveness in similar tasks~\cite{pei2016hercule}. 
This choice is further motivated by a key observation about APT behavior: attack-related activities typically form dense, interconnected communities within the provenance graph, with malicious processes exhibiting higher degrees of interaction among themselves than normal system processes, as demonstrated in~\cite{blondel2008fast}. 
By leveraging these behavioral patterns, the Louvain algorithm, combined with our temporal correlation engine, effectively identifies these related activities, revealing potential attack sequences that might otherwise remain hidden in the broader system activity.

\noindent\textbf{LLM Analyzer.}
The LLM analyzer module constitutes the third module of the \method pipeline. 
Building on the communities identified by the graph analyzer, the LLM analyzer processes the system events corresponding to the community nodes to detect malicious behaviors. 
The LLM analyzer examines the relationships and temporal sequences of operations and determines whether an attack has been observed, provides a confidence score, and maps the malicious events to the corresponding stages of the kill chain. 
When the LLM concludes that the events in the communities are highly suspicious, the corresponding nodes in the reduced graph are tagged with the confidence score; the graph is then traversed and all of the traceable nodes (processes and objects) with their corresponding edges (events) are sent to the temporal correlation engine. 

\begin{figure}[!btp]
\setlength{\abovecaptionskip}{3pt}
\setlength{\belowcaptionskip}{0pt}
\centering
\includegraphics[width=\textwidth]{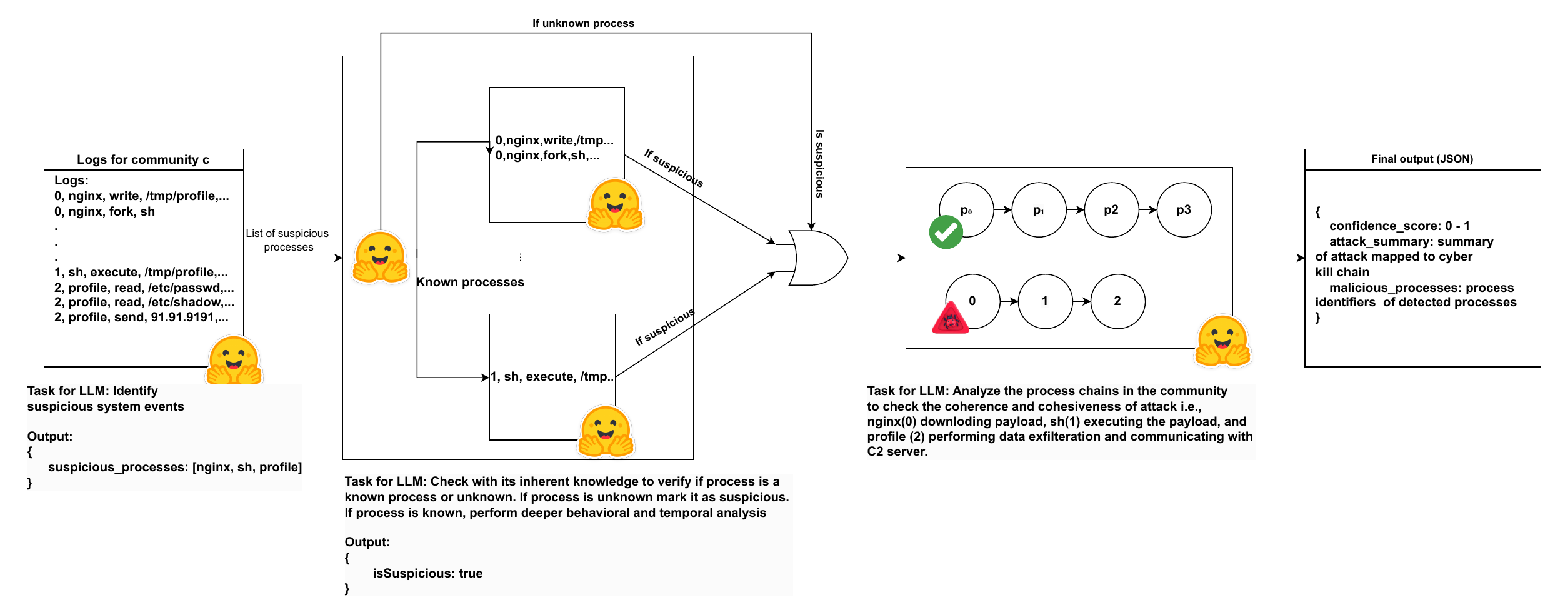}
\caption{LLM analyzer: An illustration of the CoT reasoning.}
\label{fig:cot}
\end{figure}

\noindent The methodology adopts a three-stage chain-of-thought (CoT) reasoning framework to methodically analyze complex attack scenarios.
In the initial stage, the LLM leverages its internal knowledge to identify deviations from expected system behavior patterns, such as unauthorized file writes in sensitive directories, unexpected network connections, or abnormal process spawning. 
Processes exhibiting these suspicious activities or those that are unknown to the LLM are flagged for further investigation. 
In the second stage, the LLM analyzes the temporal sequence of actions associated with each flagged process, identifying suspicious patterns that could indicate compromise (e.g., a web server process making unexpected file system modifications or establishing unusual network connections).
Finally, in the third stage, the LLM investigates inter-process relationships in a community, highlighting patterns that suggest a cohesive attack summary. 
We formalize the method in an Algorithm in Appendix~\ref{app:algorithm}. An illustration of the analysis is depicted in Fig~\ref{fig:cot}. 

\noindent Following the multi-stage analysis, the LLM assigns confidence scores. Building on prior work showing LLMs perform better with rating scales than categorization~\cite{zhuang2024}, and leveraging Freitas et al.'s use of a 0.9 confidence threshold to forward effective responses~\cite{freitas2024}, we developed a confidence scoring mechanism. Our approach assigns confidence scores ($\sigma_a$) based on detected attack sequences: $\geq$0.9 for complete attack sequences, 0.8–0.9 for partial attack sequences, and 0.7–0.8 for detected suspicious patterns.

\method sets the alert threshold ($\delta$) at 0.8 to focus analyst attention on high-confidence detections of partial or complete attacks, filtering out lower-confidence findings that might lead to alert fatigue.
When ($\sigma_a \geq \delta$), the LLM returns a three-part response. 
First, it generates a comprehensive alert for security analysts, detailing the attack description, suspicious processes, sequence of events, and their mapping to kill chain stages. 
Second, it tags nodes ($Tag(c)$), marking all suspicious processes identified by the LLM within community $c$ with their respective confidence scores. 

\noindent Finally, by traversing the graph ($Trace(c)$), it examines the reduced graph $G_r$ from the graph analyzer module, propagating confidence scores to all nodes reachable from the tagged suspicious processes. If ($\sigma_a < \delta$), the system limits its response to tagging suspicious nodes and traversing the graph.
The response can be formalized as:
\begin{equation}
	Response(c, \sigma_a) = 
	\begin{cases}
		GenerateAlert(c) + Tag(c) + Trace(c) & \text{if } \sigma_a \geq \delta, \\
		Tag(c) + Trace(c) & \text{otherwise}.
	\end{cases}
\end{equation}
Tagged and traced nodes are collected to form an attack event set $T$ in tuple format $\{\, (p_i, e_i, o_i, t_i) \,\}$ and forwarded to the correlation engine, where $p_i$ represents the process identifier, $e_i$ the event type, $o_i$ the operation, and $t_i$ the timestamp.\\

\begin{figure}[!ht]
\setlength{\abovecaptionskip}{3pt}
\setlength{\belowcaptionskip}{0pt}
\centering
\includegraphics[width=0.6\textwidth]{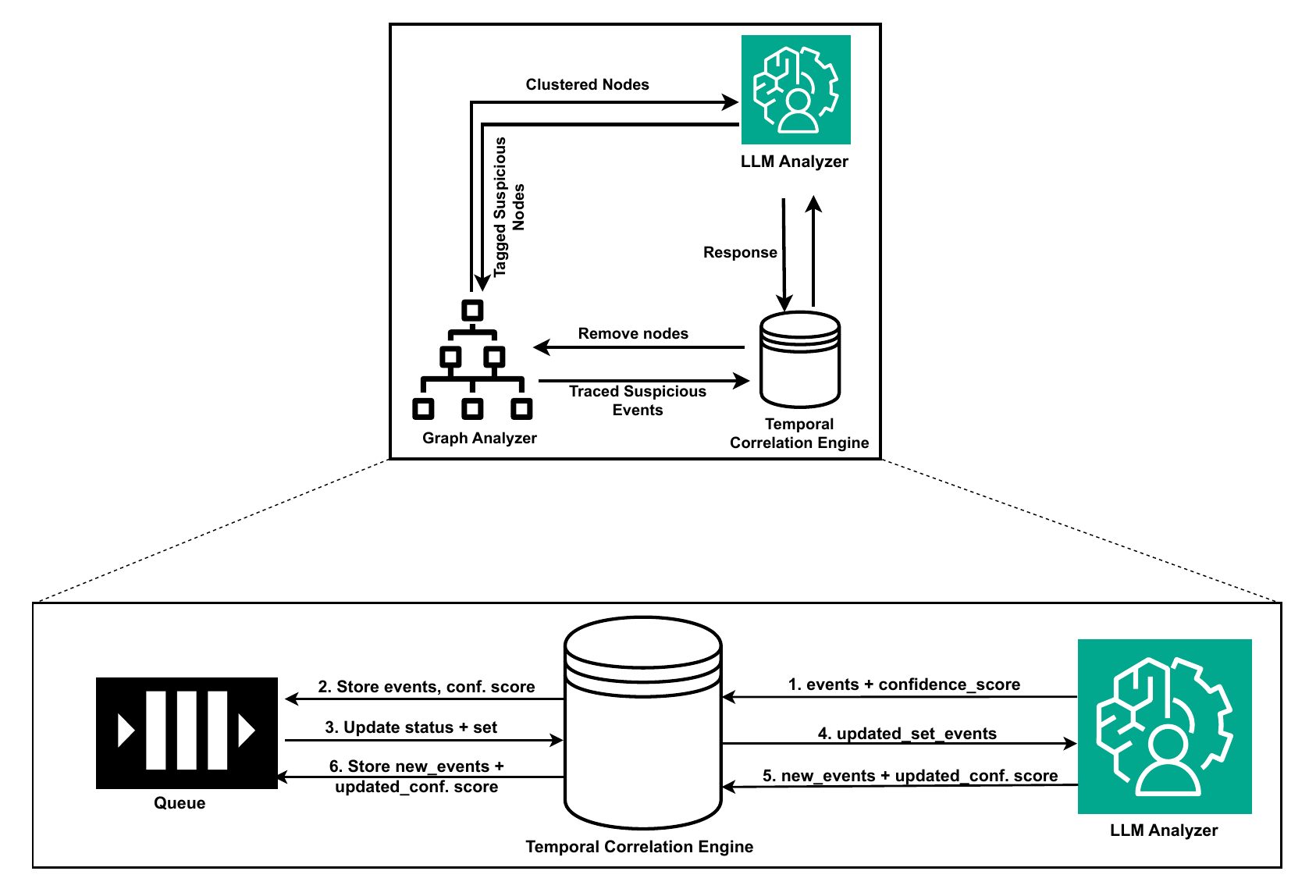}
\caption{LLM analyzer and temporal correlation engine: Analyzes system logs, tags malicious nodes, performs graph traversal, adds the tagged nodes to the attack set, and maintains historical context of attack evolution while considering memory constraints.}
\label{fig:llm_analyzer}
\end{figure}

\noindent\textbf{Temporal Correlation Engine.}
The temporal correlation engine serves as \method's orchestration module, managing both the prioritization of detected threats and the temporal aspects of long-term attack monitoring. 
It processes the attack set received from the LLM analyzer through two main mechanisms: dynamic set integration for threat assessment and rolling provenance updates for memory management.

\noindent\textit{(a) Dynamic set integration and risk stratification.} This process serves as the primary threat assessment mechanism by consolidating analysis results into a global attack set that tracks both ongoing and historical attacks. During this stage, behavioral correlation analysis leverages the confidence scores ($\sigma_a$) assigned by the LLM analyzer to identify related attack patterns, merging and filtering sets to consolidate partial APT activities into comprehensive attack chains, sorted by timestamps.
After merging, \method employs two core adjustment mechanisms that modify the initial LLM-assigned confidence scores: (1) Decay, which is applied when a tagged set continues to exhibit benign behavior over subsequent analysis windows. 
Without further suspicious activities supporting the initial classification, the confidence score gradually decreases. 
Once $\sigma_a$ falls below 0.7, the set is removed from active analysis and its nodes become eligible for pruning. (2) Reinforcement, which is activated when a set initially receives a lower confidence score (0.7 $\leq \sigma_a < 0.8$) due to early-stage attack indicators. As additional stages of the attack manifest in subsequent windows, the score increases proportionally, reflecting mounting evidence of malicious intent. This approach effectively identifies APTs that evolve slowly over time.

\noindent During set integration, \method maintains the same attack set structure $T$ and the confidence score threshold $\sigma_a$ (see LLM analyzer). 
The engine's queue management directly aligns with these thresholds: sets with $\sigma_a \geq 0.8$ enter the primary queue for immediate analyst attention, while those with $0.7 \leq \sigma_a < 0.8$ remain in the secondary queue for continued observation. 
\method continuously applies decay or reinforcement during subsequent integrations, adjusting confidence scores based on emerging evidence. 
This dual-queue prioritization process enables efficient allocation of analyst resources toward high-confidence attack chains while maintaining visibility of potentially developing threats.

\noindent\textit{(b) Rolling provenance graph updates.} \method implements a rolling provenance graph to manage memory usage when tracking long-term attacks. 
The graph continuously incorporates new process and event nodes from the system logs, with suspicious nodes being tagged with confidence scores by the LLM analyzer. 
For memory efficiency, \method removes untagged nodes after analyzing the time window $r$ and removes tagged nodes at regular intervals. 
This pruning is possible, because once suspicious activities are detected, their complete behavioral patterns and relationships are preserved in the attack sets $T$, making the actual graph nodes redundant after analysis. Through this efficient memory management technique, \method maintains comprehensive attack tracking capabilities for APTs while preventing unbounded graph growth. In Section~\ref{sec:discussion}, we present an example demonstrating how \method successfully identifies and tracks sophisticated attacks spanning multiple days.
\section{Evaluation Setup}

\noindent\textbf{Implementation.} We implemented \method in Python 3.11, using custom log parsers for data extraction. The deviation analyzer employs scikit-learn's LOF model, and the graph analyzer uses NetworkX's\cite{networkx} MultiDiGraph with the Louvain algorithm.\footnote{\url{https://github.com/taynaud/python-louvain}}
We developed a custom filtering mechanism to filter suspicious file operations, network communications, and process-file operations. For the LLM, we deployed Qwen 2.5 (32B) in 8-bit quantized mode on an NVIDIA RTX 6000 GPU (32GB VRAM). \\

\noindent\textbf{Datasets}
Our evaluation leverages three system-level audit log datasets (THEIA, CADETS, Public Arena) and a high-level application log dataset (BlindEagle) in diverse environments (in Table~\ref{tab:train_test} we provide information on the datasets): 

\begin{itemize}[leftmargin=*,noitemsep]
    \item \textit{DARPA eng.3 THEIA:} A comprehensive Ubuntu Linux dataset comprising 106 million system audit logs (40GB) collected over 12 days. 
    It contains two stages of a successful APT attack instance and one failed attempt, with the successful attack spanning three days—(ideal for evaluating long-duration, multi-stage APT detection capabilities).

    \item \textit{DARPA eng.3 CADETS:} A FreeBSD-based dataset containing 42 million audit logs (25GB) captured over 12 days, featuring four APT attack instances (three successful and one failed). This dataset complements THEIA by providing attack patterns in a different Unix-like environment.

    \item \textit{Public Arena:} A Windows-focused dataset collected from two hosts in a simulated public cloud environment over six days. It contains one instance of an APT attack. With 16 million audit records (7GB), it enables cross-platform validation of our framework. 

    \item \textit{High-Level Application Log:}
To validate \method's applicability in enterprise environments, we developed an in-house testbed emulating the Blind Eagle APT campaign. 
The experiment ran for 60 minutes, generating 6,100 Splunk logs across three Windows workstations running Windows 10 and Windows 11. 
We simulated the attack by compromising two out of the workstations while incorporating realistic background user activities, such as web browsing, document editing, and email usage, in our controlled environment. 
This setup enabled us to evaluate \method's efficacy in detecting APT patterns and distinguishing them from benign user behaviors.
\end{itemize}

To further ensure that our evaluation reflects real-world scenarios, we split each dataset into training and testing windows. 
For training, we used data from earlier logs, prior to the occurrence of the first attack event, minimizing data leakage and simulating the challenges of deploying \method in a real-time environment (see Table~\ref{tab:train_test} for the time windows). 
For training, we used 28\% of the CADETS logs, 35\% of the THEIA logs, 29\% of the Public Arena logs and 33\% of the Blind Eagle APT logs, respectively. 
ensures both consistency across datasets and that there is sufficient historical context for extended inference periods during testing.

DARPA released the ground truths for CADETS and THEIA datasets containing IoCs but not specific attack events. 
To find the relevant attack events, we developed a customized script to retrieve logs containing these IoCs by cross-referencing log timestamps with the attack times delineated in the ground truth. 
We then manually annotated each attack event from these logs. 
For the Public Arena and Blind Eagle datasets we already had the ground truths with the identified attack events.
Table~\ref{tab:train_test} summarizes the characteristics the datasets, including key provenance statistics (e.g., number of edges, nodes, malicious entities), and overall dataset size. 
These datasets and their ground truth mappings served as the foundation for our comprehensive evaluation methodology.

\begin{table}[!ht]
\centering
\caption{Train and test windows with detailed provenance statistics.}

\begin{adjustbox}{width=1.0\textwidth,center=\textwidth}
\begin{tabular}{|l|l|c|c|c|c|c|c|c|c|}
\hline
\multirow{2}{*}{\textbf{Dataset}} & \multirow{2}{*}{\textbf{Split Time}} & \multicolumn{2}{c|}{\textbf{Train}} & \multicolumn{5}{c|}{\textbf{Test}} \\
\cline{3-9}
 &  & \textbf{Duration} & \textbf{\# Logs} & \textbf{Duration} & \textbf{\# Logs} & \textbf{\# Attacks} & \multicolumn{2}{c|}{\textbf{Provenance Statistics}} \\
\cline{8-9}
 & & & & & & & \textbf{Benign (N, E, KE)} & \textbf{Malicious (N, E, KE)} \\
\hline
CADETS & 2018-04-06 11:00:00 & 6D8H & 4,239,474 & 3D4H & 10,949,668 & 3 & 263,775, 10,947,794, 1,293,534 & 33, 2,037, 60 \\

THEIA & 2018-04-09 22:15:12 & 3D13H & 9,654,772 & 7D11H & 17,827,942 & 1 & 492,556, 17,827,833, 1,823,963 & 16, 170, 31 \\

Public Arena & 2022-05-13 00:00:00 & 3D9H & 2,593,769 & 7D9H & 6,468,573 & 1 & 25,527, 6,093,093, 106,285 & 6, 375,480, 23 \\

Blind Eagle & 2024-11-06 16:18:00 & 20M & 700 & 40M & 6149 & 2 & 651, 5886, 4354 & 37, 263, 114 \\
\hline
\end{tabular}
\end{adjustbox}
\textit{N - nodes, E - edges, KE - key edges;} \\
\textit{*High due to repetitive communication between the malicious nodes.}

\label{tab:train_test}
\end{table}

\noindent\textbf{Evaluation Metrics.}
To evaluate \method's performance, we employed a two-phase analysis process: coarse-grained (time-window level) detection and fine-grained (event-level) detection. 
This dual-layer approach enables comprehensive evaluation of the system's detection capabilities.
We used standard performance metrics: precision, recall, and the F1-score. 
At the event level, we focused on measuring the alert quality and assess \method's effectiveness in reducing alert fatigue, a critical consideration for practical deployment in security operations.
We compare our framework's performance to that of state-of-the-art APT detection systems, including UNICORN~\cite{han2020unicorn}, DeepLog~\cite{du2017deeplog}, and KAIROS~\cite{cheng2023kairospracticalintrusiondetection}
For fairness, we used the original implementations and hyperparameters provided by the respective authors. 
UNICORN and KAIROS were adapted to our sliding window mechanism, ensuring consistent evaluation across methods while preserving their core functionality.
\section{Evaluation Results}\label{sec:results}

\noindent\textbf{Performance Analysis of Deviation Analyzer.}
We start by evaluating the performance of the deviation analyzer, focusing on its detection capabilities and log reduction efficiency. 
The deviation analyzer demonstrates exceptional detection capabilities with 100\% recall on all datasets, successfully identifying every known attack event while maintaining a minimum log reduction of approximately 20 times across all datasets. 
Its adaptive thresholding mechanism (i.e., the contamination threshold) effectively distinguishes malicious patterns from benign system behaviors. 
The analyzer's efficiency in processing large-scale log data with minimal computational overhead makes it particularly valuable for real-time APT detection in enterprise environments. \\

\noindent\textbf{Performance Analysis of Graph Analyzer. }
Next, we examine the graph analyzer's effectiveness in preserving attack events while while significantly reducing the number of logs on diverse datasets. 
The graph analyzer demonstrates effective capabilities in preserving attack events while achieving significant log reduction on all datasets. 
We achieved a mean reduction of 95.58\% across the datasets, with standard deviations ranging from 1.98\% to 6.59\%, substantially reducing the computational overhead in analysis. 
In terms of recall, the analyzer maintains near-perfect attack event retention across datasets with rates consistently exceeding 99\%. 
This exceptional performance is demonstrated by minimal attack event loss - only three events across millions of log entries - showcasing the module's ability to preserve critical event information even in complex environments with diverse log patterns. 
These results suggest significant practical implications for large-scale security monitoring systems, where it can substantially reduce storage and processing requirements while maintaining comprehensive threat detection capabilities. \\

\noindent\textbf{Time Window-Based Performance Analysis of \method. }
Having established the effectiveness of two of the modules, we  evaluate \method's overall performance using a systematic time window-based approach. 
We employed a sliding window mechanism with 30-minute intervals and a 15-minute step size. 
This allowed for continuous monitoring and evaluation of system behavior. 
The CADETS dataset was divided into 714 windows, of which only 12 contained attack sequences, illustrating the imbalanced nature of the data. 
Similarly, the THEIA dataset comprised 303 windows, with nine containing attacks, reflecting a similar skew.
In addition, in this initial evaluation phase, we assessed several local language models. 
Among the candidates—Qwen 2.5, 
EXAONE 3.5, and LLama 3.1—Qwen 2.5 demonstrated superior and consistent performance on all datasets 
In Appendix ~\ref{app:model_eval} we provide detailed results of our comparative study for model selection. \\

\noindent\textbf{Event-Based Performance Analysis of \method. } 
While the window-based metrics provide a high-level view of \method's detection capabilities, a more granular event-based analysis is necessary to evaluate its practical utility for security operations. \method's primary objective is to alleviate the workload of security analysts. Instead of requiring analysts to perform extensive triage on large volumes of false positive alerts, \method automatically summarizes what it deems to be an attack, enabling analysts to quickly evaluate whether malicious activity is present in the system. Even in scenarios where alerts might be false positives, the consolidated summary containing IoCs can help analysts decide if further action is necessary. Since \method bases its summaries on identified attack events, it is critical to examine these events to understand the extent and accuracy of \method's detection capabilities.

\noindent Table~\ref{tab:attack_evaluation} highlights \method's capabilities on four diverse datasets. On the CADETS dataset, \method consistently demonstrates perfect precision coupled with high recall for all attacks. The method achieves perfect recall on the THEIA and Public Arena datasets, effectively detecting all attack events. For Blind Eagle, \method maintains high recall, underscoring its effectiveness in diverse environments and threat scenarios.

\begin{table}[ht]
\centering
\caption{Reduced Graph Metrics of \method: precision, recall, TP, FP, and FN across CADETS, THEIA, Public Arena, and Blind Eagle datasets.}
\label{tab:attack_evaluation}
\adjustbox{max width=0.99\textwidth}{%
\begin{tabular}{|l|c|c|c|c|c|c|}
\hline
\textbf{Dataset} & \textbf{Attack} & \textbf{TP} & \textbf{FP} & \textbf{FN} & \textbf{Precision} & \textbf{Recall} \\ 
\hline
\multirow{3}{*}{CADETS} 
    & Attack 1 & 13 & 0 & 1 & 1.00 & 0.93 \\ \cline{2-7}
    & Attack 2 & 25 & 0 & 2 & 1.00 & 0.93 \\ \cline{2-7}
    & Attack 3 & 19 & 0 & 1 & 1.00 & 0.95 \\ 
\hline
THEIA & Attack 1 & 32 & 47 & 0 & 0.40 & 1.00 \\ 
\hline
Public Arena & Attack 1 & 23 & 251 & 0 & 0.08 & 1.00 \\ 
\hline
Blind Eagle & Attack 1 & 31 & 35 & 5 & 0.46 & 0.86 \\ 
\hline
\end{tabular}
}
\end{table}

\noindent\textbf{Comparative Analysis.} We compare \method's performance for time-window based detection against state-of-the-art detection systems, including KAIROS, UNICORN, and DeepLog, on both the CADETS and THEIA datasets by adapting them to our time-window based approach. While Public Arena's parsers were unavailable and Blind Eagle's high-level log format was incompatible with provenance-based Intrusion Detection System, the comprehensive comparison on compatible datasets demonstrates \method's competitive performance on multiple key metrics, as summarized in Table~\ref{tab:performance-comparison}.

\begin{table}[ht]
\caption{Comparison of time-window based detection performance between  different methods on CADETS and THEIA datasets.}
\label{tab:performance-comparison}
\centering
\adjustbox{max width=0.99\textwidth}{%
\begin{tabular}{|l|l|c|cccc|ccc|}
\hline
\multirow{2}{*}{\textbf{Method}} & \multirow{2}{*}{\textbf{Dataset}} & 
\multirow{2}{*}{\textbf{Attack Identified}} & 
\multicolumn{4}{c|}{\textbf{Confusion Matrix}} & 
\multicolumn{3}{c|}{\textbf{Performance Metrics}} \\
\cline{4-10}
 & & & \textbf{TP} & \textbf{FN} & \textbf{FP} & \textbf{TN} & \textbf{Precision} & \textbf{Recall} & \textbf{F1-Score} \\
\hline
\multirow{2}{*}{SHIELD} 
  & CADETS & 3/3 & 7 & 5 & 17 & 685 & 0.29 & 0.58 & 0.39 \\
  & THEIA  & 2/2 & 6 & 3 & 4  & 290 & 0.60 & 0.67 & 0.63 \\
\hline
\multirow{2}{*}{KAIROS} 
  & CADETS & 2/3 & 9 & 3 & 10 & 692 & 0.64 & 0.75 & 0.69 \\
  & THEIA  & 2/2 & 5 & 4 & 11 & 283 & 0.31 & 0.55 & 0.40 \\
\hline
\multirow{2}{*}{Unicorn} 
  & CADETS & 3/3 & 4 & 8 & 28 & 674 & 0.13 & 0.33 & 0.18 \\
  & THEIA  & 2/2 & 4 & 5 & 23 & 271 & 0.15 & 0.44 & 0.22 \\
\hline
\multirow{2}{*}{DeepLog} 
  & CADETS & 3/3 & 9 & 3 & 683 & 19 & 0.01 & 0.75 & 0.03 \\
  & THEIA  & 2/2 & 4 & 5 & 60 & 234 & 0.06 & 0.44 & 0.11 \\
\hline
\end{tabular}
}
\end{table}

On the  CADETS dataset, \method successfully identified all attacks while maintaining balanced precision and recall metrics. Although KAIROS exhibited higher precision and recall, it failed to detect one attack, revealing a critical gap in coverage. UNICORN and DeepLog, while detecting all attacks, suffer from higher false positive rates. On THEIA, \method demonstrated superior performance, achieving the highest F1 Score among all evaluated approaches.

To further benchmark \method, we evaluated its event-based detection capabilities against other methods. Table~\ref{tab:method_comparison} presents a detailed comparison of \method, KAIROS, UNICORN, and DeepLog on the CADETS dataset. The results demonstrate \method's superior precision, maintaining perfect precision across all attacks while achieving consistently high recall. KAIROS exhibited mixed performance, achieving near-perfect recall but suffering from low precision and critically failing to detect an attack. In contrast, while UNICORN and DeepLog achieved high recall, they generated significant false positive alarms. \\

\begin{table}[!ht]
\centering
\caption{Comparison of event-level detection performance between  different methods on CADETS dataset.}
\label{tab:method_comparison}
\adjustbox{max width=0.99\textwidth}{%
\begin{tabular}{|l|c|c|c|c|c|c|}
\hline
\textbf{Method} & \textbf{Attack} & \textbf{TP} & \textbf{FP} & \textbf{FN} & \textbf{Precision} & \textbf{Recall} \\ 
\hline
\multirow{3}{*}{SHIELD} 
    & Attack 1 & 13 & 0 & 1 & 1.00 & 0.93 \\ \cline{2-7} 
    & Attack 2 & 25 & 0 & 2 & 1.00 & 0.93 \\ \cline{2-7}
    & Attack 3 & 19 & 0 & 1 & 1.00 & 0.95 \\ 
\hline
\multirow{3}{*}{KAIROS}
    & Attack 1 & 14 & 108 & 0 & 0.11 & 1.00 \\ \cline{2-7}
    & Attack 2 & 25 & 129 & 2 & 0.16 & 0.93 \\ \cline{2-7}
    & Attack 3 & \multicolumn{5}{c|}{Not detected} \\ 
\hline
\multirow{3}{*}{UNICORN}
    & Attack 1 & 14 & 3539 & 0 & 0.004 & 1.00 \\ \cline{2-7}
    & Attack 2 & 27 & 4561 & 0 & 0.006 & 1.00 \\ \cline{2-7}
    & Attack 3 & 20 & 3978 & 0 & 0.005 & 1.00 \\ 
\hline
\multirow{3}{*}{DeepLog}
    & Attack 1 & 6 & 3543 & 4 & 0.003 & 0.71 \\ \cline{2-7}
    & Attack 2 & 25 & 4561 & 2 & 0.005 & 0.93 \\ \cline{2-7}
    & Attack 3 & 11 & 3982 & 5 & 0.003 & 0.55 \\ 
\hline
\end{tabular}
}
\end{table}

\noindent\textbf{An Illustration of the Attack Summary Quality. }
While quantitative metrics (precision, recall etc.) demonstrate \method's detection efficacy, the system's ability to generate interpretable attack summaries is equally crucial. 
\method exhibits exceptional capabilities in generating comprehensive attack summaries that align with the cyber kill-chain framework. 
In Appendix~\ref{app:attack_summary}, we present the summary generated by \method for the THEIA dataset, which provides a detailed visualization of the complete attack progression. During the delivery phase, the system precisely identified shell code servers (\texttt{61.130.69.232} and \texttt{141.43.176.203}). It then accurately traced the exploitation chain through the firefox, clean, and profile processes. The analysis captured the installation of persistence mechanisms via \texttt{/etc/firefox/native-messaging-hosts/gtcache} and effectively tracked command and control (C2) activities. Additionally, \method identified potential data exfiltration patterns through the profile and mail processes, demonstrating its capability to trace the attack's impact comprehensively.
The aforementioned results substantiate \method's efficacy in generating detailed and structured attack summaries, accurately identifying crucial attack components across various stages of the kill chain.
\section{Discussion}\label{sec:discussion}
\noindent\textbf{Attack Detection Effectiveness.} The effectiveness of our reinforcement and decay mechanism is demonstrated in both attack detection and false positive reduction. 
For instance, in the THEIA dataset, a process called \textit{profile} performs C2 activity on day one and then goes dormant. 
On day three, it downloads a malicious payload into \textit{/var/log/mail} and forks a process called \textit{mail} which performs further malicious activities. 
The LLM initially identifies the exploitation from \textit{firefox} to \textit{profile} on day one, marking it as a partial attack with a probability score of 0.85 and placing it in a set. 
When \texttt{profile}'s activity on day three is detected, it is also marked as a partial attack with a score of 0.85 and added to another set. 
The temporal correlation engine maps these activities to the same set, and upon merging these sets, it reassigns the new set to the LLM for analysis. The LLM reinforces the confidence score to 0.91, indicating that it identified the complete attack.

\noindent On the CADETS dataset, \method demonstrates its decay mechanism's effectiveness in handling false positives. 
When analyzing activities for process chain containing \textit{imapd}, \textit{wget}, and \textit{links}, the LLM initially assigns confidence score around 0.85 due to potentially suspicious activities. 
However, as these processes exhibit benign operations in subsequent analysis windows, the temporal correlation engine periodically forwards these observations to the LLM. 
Observing no suspicious progression the LLM gradually decreases the confidence score to 0.75. 
Due to the reduction in score, the engine transfers these processes to a secondary monitoring queue, temporarily suspending alert generation. 
This decay mechanism effectively addresses alert fatigue in security operations by ensuring that analysts focus on high-confidence threats. \\

\noindent\textbf{Zero-Day Attack Detection Capabilities.} Our evaluation on four datasets, including real-world and benchmark scenarios, demonstrated \method's effectiveness in detecting previously unknown attacks. 
Since attackers must progress through certain essential stages regardless of their specific techniques, \method can identify suspicious patterns at the kernel level even when confronted with novel attack methods.
\method's success in zero-day detection relies on two key factors: the invariant nature of kernel-level attack traces and the LLM's comprehensive understanding of system behavior patterns. 
While attack implementations may vary, the underlying progression through attack stages creates detectable patterns that \method can recognize without prior exposure to specific variants. 
Through these capabilities, \method demonstrates a significant advancement in automated attack detection and investigation.
The framework's ability to correlate events across extended time periods, manage false positives, and detect novel attack patterns makes it as a valuable tool for modern security operations. \\

\noindent\textbf{Limitations.}
\method employs a 15-minute sliding window for attack detection, introducing a maximum detection delay of 15 minutes in edge cases.
However, this delay remains well within acceptable operational parameters, as it significantly outperforms typical SOC mean time to resolve metrics in real-world deployment.\footnote{\url{https://www.ibm.com/reports/
data-breach}}
In addition, 
the multi-stage pipeline, while enabling comprehensive threat detection, presents sizeable computational demands. 
The deployment of the deviation and graph analyzer modules on endpoint systems contributes to resource utilization on those devices. 
Additionally, the local deployment of our chosen LLM requires substantial GPU resources (32GB VRAM), which may limit deployment options in resource-constrained environments.

\section{Conclusions and Future Work}

In this paper, we presented \method, an innovative framework that combines anomaly detection and graph analysis with LLM reasoning for attack detection, investigation, and contextual analysis.
Future work can explore integrating automated incident response capabilities using contextual insights and addressing the challenges of parsing high-level logs, as demonstrated in our Blind Eagle proof-of-concept. 
While our approach effectively analyzes high-level logs, its success depends on complete log capture, highlighting the need to overcome challenges like logging misconfigurations or incomplete capture.

\bibliographystyle{splncs04}
\bibliography{references}

\appendix

\section{Chain-of-Thought Detection Algorithm}
\label{app:algorithm}

\begin{algorithm}[h!]
  \small
  \caption{Chain-of-Thought detection}
  \label{alg:chain_prompting}
  \begin{algorithmic}[1]
  
    \REQUIRE Communities $C$, LLM model $M$
    \ENSURE Attack alerts, summaries, tagged processes
    \FOR{each community $c \in C$}
      \STATE $logs \gets$ GetLogsFromNodes$(c)$
      \STATE $suspiciousProcesses \gets$ CheckKnownBehavior$(logs, M)$
      \IF{$|suspiciousProcesses|$}
        \STATE $deviationFound \gets$ \textbf{False}
        \STATE $temporalPatterns \gets \emptyset$
        \FOR{each process $p \in suspiciousProcesses$}
          \STATE $processLogs \gets$ GetProcessLogs$(p)$
          \STATE $deviation \gets$ AnalyzeBehavior$(processLogs, M)$
          \IF{$deviation \neq \emptyset$}
            \STATE $deviationFound \gets$ \textbf{True}
            \STATE $temporalPatterns[p] \gets deviation$
          \ENDIF
        \ENDFOR
        \IF{$deviationFound$}
          \STATE $attackChain \gets$ ExtractProcessChain$(suspiciousProcesses, temporalPatterns, logs)$
          \IF{$attackChain \neq \emptyset$}
            \STATE $(score, summary, tagged) \gets$ AnalyzeChain$(attackChain, M)$
            \IF{$score \geq 0.8$}
              \RETURN RaiseAlert$()$, $summary$, $tagged$, $score$
            \ELSIF{$0.7 \leq score < 0.8$}
              \RETURN $tagged$, $score$
            \ENDIF
          \ENDIF
        \ENDIF
      \ENDIF
    \ENDFOR
    \RETURN $\emptyset$
  \end{algorithmic}
\end{algorithm}
\clearpage

\section{Comparative Study of LLM Models}
\label{app:model_eval}

\begin{table}[h]
\centering
\caption{Performance comparison of different models across datasets.}
\label{tab:performance}
\adjustbox{max width=0.8\textwidth}{%
\begin{tabular}{|c|c|cccc|ccc|}
\hline
\textbf{Model} & \textbf{Dataset} &
  \multicolumn{4}{c|}{\textbf{Confusion Matrix}} &
  \multicolumn{3}{c|}{\textbf{Performance Metrics}} \\
\cline{3-9}
 & & \textbf{TP} & \textbf{FN} & \textbf{FP} & \textbf{TN} &
   \textbf{Precision} & \textbf{Recall} & \textbf{F1-Score} \\
\hline
\multirow{4}{*}{GPT-4o}
  & CADETS     & 07 & 05 & 21 & 681 & 25.0  & 58.3  & 35.0 \\
  & THEIA      & 06 & 03 & 00 & 294 & 100.0 & 66.7  & 80.0 \\
  & Arena      & 03 & 00 & 00 & 235 & 100.0 & 100.0 & 100.0 \\
  & BlindEagle & 01 & 00 & 00 & 00  & 100.0 & 100.0 & 100.0 \\
\hline
\multirow{4}{*}{Qwen-2.5}
  & CADETS     & 07 & 05 & 17 & 685 & 29.2  & 58.3  & 38.9 \\
  & THEIA      & 06 & 03 & 04 & 290 & 60.0  & 66.7  & 63.2 \\
  & Arena      & 03 & 00 & 00 & 235 & 100.0 & 100.0 & 100.0 \\
  & BlindEagle & 01 & 00 & 00 & 00  & 100.0 & 100.0 & 100.0 \\
\hline
\multirow{4}{*}{EXAONE-3.5}
  & CADETS     & 07 & 05 & 31 & 671 & 18.4  & 58.3  & 28.0 \\
  & THEIA      & 04 & 05 & 00 & 294 & 100.0 & 44.4  & 61.5 \\
  & Arena      & 02 & 01 & 00 & 235 & 100.0 & 67.0  & 80.2 \\
  & BlindEagle & 01 & 00 & 00 & 00  & 100.0 & 100.0 & 100.0 \\
\hline
\multirow{4}{*}{LLama-3.1}
  & CADETS     & 06 & 06 & 48 & 654 & 11.1  & 50.0  & 18.2 \\
  & THEIA      & 05 & 04 & 03 & 291 & 62.5  & 55.6  & 58.8 \\
  & Arena      & 03 & 00 & 02 & 233 & 60.0  & 100.0 & 75.0 \\
  & BlindEagle & 01 & 00 & 00 & 00  & 100.0 & 100.0 & 100.0 \\
\hline
\end{tabular}
}
\end{table}

\section{Attack Summary Visualization}
\label{app:attack_summary}
\begin{figure}[h!]
    \makebox[\textwidth][c]{
        \includegraphics[width=\textwidth]{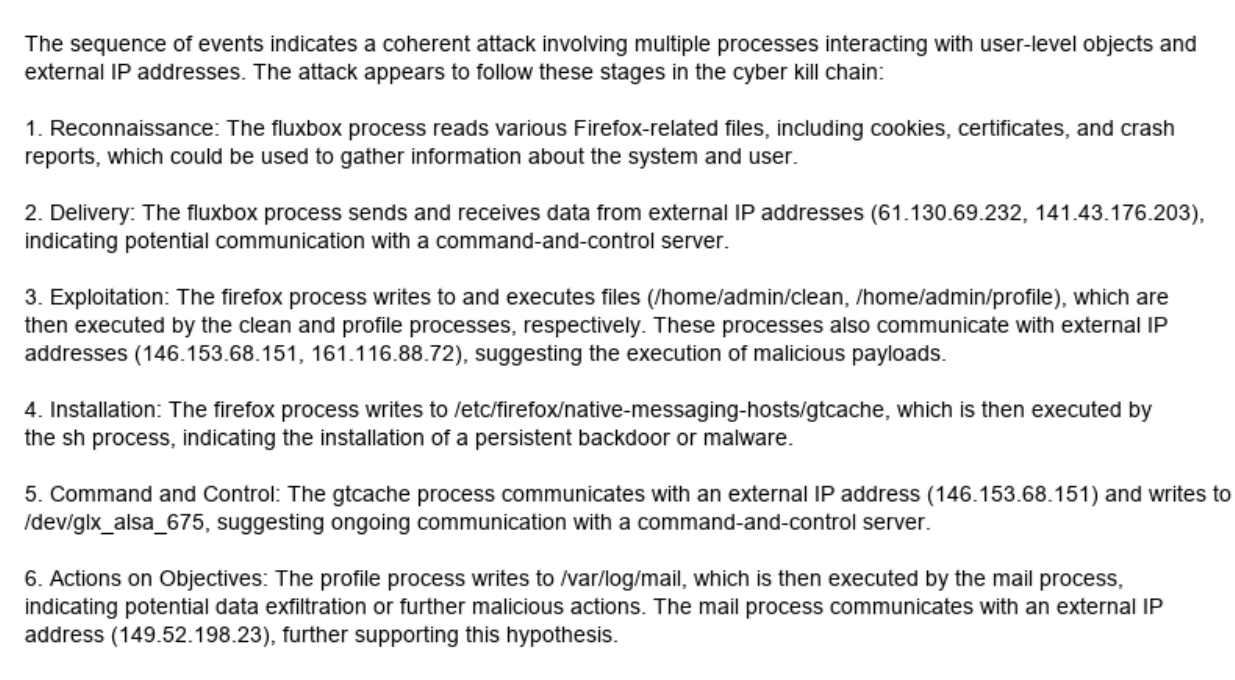}
    }
    \caption{An illustration of attack summary generated for the attack in THEIA dataset.}
    \label{fig:attacksum}
\end{figure}

\end{document}